# Non-equilibrium phase precursors to the insulator-metal transition in V$_2$O$_3$

Andrej Singer[1,*], Juan Gabriel Ramirez[1,2], Ilya Valmianski[1], Devin Cela[1], Nelson Hua[1], Roopali Kukreja[1], James Wingert[1], Olesya Kovalchuk[1], James M. Glownia[3], Marcin Sikiroski[3], Matthieu Chollet[3], Martin Holt[4], Ivan K. Schuller[1], and Oleg G. Shpyrko[1]

[1]*Department of Physics and Center for Advanced Nanoscience, University of California San Diego, La Jolla, California 92093, USA*

[2] *Department of Physics, Universidad de los Andes, Bogotá 111711, Colombia*

[3]*LCLS, SLAC National Accelerator Laboratory, Menlo Park, California 94025, USA*

[4]*Center for Nanoscale Materials, Argonne National Laboratory, Argonne, IL 60439, USA*

[*]*Present address: Department of Materials Science and Engineering, Cornell University, Ithaca, NY, 14850, USA*

**The discovery of novel phases of matter is at the core of modern physics. In quantum materials, subtle variations in atomic-scale interactions can induce dramatic changes in macroscopic properties and drive phase transitions[1,2]. Despite their importance[3–6], the mesoscale processes underpinning phase transitions often remain elusive because of the vast differences in timescales between atomic and electronic changes and thermodynamic transformations. Here, we photoinduce and directly observe with x-ray scattering an ultrafast enhancement of the structural long-range order in the archetypal Mott system V$_2$O$_3$. Despite the ultrafast change in crystal symmetry, the change of unit cell volume occurs an order of magnitude slower and coincides with the insulator-to-metal transition[7]. The decoupling between the two structural responses in the time domain highlights the existence of a transient photoinduced precursor phase, which is distinct from the two structural phases present in equilibrium. X-ray nanoscopy reveals that acoustic phonons trapped in nanoscale blocks govern the dynamics of the ultrafast transition into the precursor phase, while nucleation and growth of metallic domains dictate the duration of the slower transition into the metallic phase[7]. The enhancement of the long-range order before completion of the electronic transition demonstrates the critical role the non-equilibrium structural phases play during electronic phase transitions in correlated electrons systems.**

Vanadium sesquioxide (V$_2$O$_3$) undergoes an insulator-to-metal transition upon heating at 160 K: the electric conductivity grows by five orders of magnitude (see Fig. S1) via percolation of the high-temperature (HT) paramagnetic, metallic domains within the low-temperature (LT) antiferromagnetic, insulating domains [8–10]. The electronic phase transition is accompanied by a structural phase transition, where the crystal symmetry increases from monoclinic to rhombohedral, and the volume shrinks via reduction of the average hexagon edge from 2.91 Å to 2.87 Å (see Fig. 1**a**)[11–13]. While the symmetry and volume are coupled in equilibrium, here we show that ultrafast photoexcitation decouples these structural degrees of freedom, and the phase transformation occurs via a non-equilibrium transient pathway (see Fig. 1**b**).

We study a 100 nm thin V$_2$O$_3$ film grown along the (024)$_{Rh}$ crystallographic direction (see Supplement). The difference in the hexagon edge lengths results in two well distinct X-ray Bragg peaks along the film normal, $q_\perp$ : (024)$_{Rh}$ and (022)$_{Mon}$ (see inset in Fig. 1**c**). The LT diffraction peak is broader than the HT peak in the direction parallel to the film, $q_\parallel$ (see inset in Fig. 1**c**). The excess broadening results from a reduction of the long-range order (coherence length) due to the breaking of the crystal symmetry at low temperatures: three monoclinic structures are possible, each distorted along a different hexagon edge[11,12]. When heated quasi-statically through the phase transition temperature, the height of the HT peak grows, while the height of the LT peak decreases monotonically. The LT diffraction peak width is proportional to the LT peak height, while the width of the HT peak is static (see Fig. S2).

We use short, optical laser pulses (E=1.55 eV, 40 fs duration, 500 μm spot size) to induce an insulator-to-metal phase transition in V$_2$O$_3$. We probe the structural response to photoexcitation with short x-ray pulses (E=9 keV, 10 fs duration, 200 μm spot size) in stroboscopic mode at the Linac Coherent Light Source[14,15]. After photoexcitation, the height of the HT peak grows as expected (see Fig. 1**c**), as the optical laser pulse heats the film. Because the HT peak grows, one may expect a drop in the LT peak height. Unexpectedly, the height of the LT peak grows for 2.5 ps before it declines for larger time delays (see Fig. 1**d**). From the 4D data set (3D reciprocal space and time, see

Supplement and Fig. S3), we find that the LT peak height grows by narrowing the LT peak. Since the peak broadening in the ground state originates from the decrease in long range order, we conclude that photoexcitation induces higher symmetry in the LT phase after 2.5 ps. At these short timescales however, the photoexcited LT phase retains its unit cell volume, as revealed by the unmodified Bragg peak position (see Fig. 1**b**).

The photoinduced structural phase transition displays two processes with vastly different characteristic time scales. First, the structural coherence length in the photoexcited LT phase, $\xi(\tau)$, grows rapidly within 2.5 ps (see Fig 2**a**, left axis) and remains high for more than 100 ps. Second, the HT phase fraction, $P_{HT}(\tau)$, grows gradually within 100 ps after photoexcitation. The unit cell volume of the photoexcited LT regions shrinks discontinuously and the HT phase emerges, evidenced by decrease of the LT and increase of the HT integral peak intensity (see right axis in Fig. 2**a**, and Fig. S4). Furthermore, the coherence length and the phase fraction depend differently on the pump fluence. The coherence length $\xi(2.5ps)$ increases linearly with the pump fluence before it saturates at ~5 mJ/cm$^2$ (see Fig. 2**b**). The dynamic saturation value at 1.3 above the ground state agrees with the maximum coherence length of the LT phase measured in equilibrium (see Fig. S2). In contrast, the final HT phase fraction $P_{HT}(100\ ps)$ displays a fluence threshold[7,16] around 1 mJ/cm$^2$: the growth rate increases significantly when the pump energy overcomes the energy barrier due to the latent heat (see Fig. 2**c**). At fluences higher than 5 mJ/cm$^2$ the phase fraction $P_{HT}(100\ ps)$ saturates at its maximum value of one, when the entire volume is in the HT phase. While the presence of the fluence threshold and the duration of the HT phase growth are both consistent with the photoinduced electronic transition[7], the origin of the 2.5 ps time scale remains elusive. To understand the mechanism underpinning this ultrafast coherence length growth, we use X-ray nanoscopy.

We map the static nanoscale spatial distribution of the two structural phases (see Fig. 3**a**) and their coherence length (see Fig. S5) by scanning a 20 nm X-ray beam across the sample. The typical size of an LT domain (connected blue region in Fig. 3**a**) is 100 nm, consistent with near-field imaging of insulating and metallic pockets[10]. Surprisingly, the broadening of the LT diffraction peak is present in the diffraction collected during a single exposure by the 20nm nanobeam (see Fig. 3**b**). Thus, the monoclinic distortion reduces the long-range order on a length scale much smaller than the size of a LT domain. X-ray diffraction at higher angles (see Fig. S6) reveals splitting of the LT phase into a mosaic of two distinct types of crystallites angularly misaligned to each other by 1° about the film normal. The two categories correspond to monoclinic distortions either along the A or B directions (see Figs. 3**c** and Fig. S7). Statistical analysis of 20,000 diffraction patterns collected while scanning the nanobeam across the film (see Supplement and Fig. S8) yields an average crystallite size of $\xi$=8 nm (see Fig. 3**c**). The spatially resolved peak broadening (see Fig. S5b) is independent of the LT peak intensity, suggesting a random distribution of the crystallites throughout the film. The diffraction peak corresponding to the monoclinic distortion along the third direction appears at a different momentum transfer[11] and is only noticeable at the lowest temperatures.

The combination of time-resolved diffraction and x-ray nanoscopy implies that the photoexcited structural phase transition proceeds as follows (see Fig. 3**d**). Photoexcitation enhances the coherence length through a collective alignment of the crystallites within 2.5 ps. This ultrafast time scale (see Fig. 1**d**) is in agreement with the half period $\frac{\tau_P}{2} = \frac{L}{v} \approx 3.2\ ps$ of the coherent phonon at the Brillouin zone boundary[17], where L=8 nm is the size of the crystallite and $v = 2.5$ nm/ps is the sound velocity in monoclinic $V_2O_3$ (Ref. [18]). The 8 nm large crystallites act as internal traps for acoustic phonons, which drive the symmetry enhancement by shear motion[19]. The degree of alignment grows linearly with pump fluence (see Fig. 2**b**), characteristic of a continuous phase transition as described by Landau theory extended to ultrafast transitions[20]. The photoinduced phase transition completes with a unit cell volume contraction within 100 ps, while the crystallites remain aligned with an enhanced coherence length. This longer process displays a fluence threshold: the hallmark of a first-order phase transition due to the latent heat [7,16]. The volume contraction is governed by the nucleation and ballistic growth through 100 nm large LT domains moving at the sound velocity [7] (see Fig. 3**a**).

Laser pulse energies well below the fluence threshold align the crystallites in the LT phase, yet avoid the transition into the HT phase (see Fig. S9). At these low fluences the enhanced long-range order prevails for more than 100 ps as measured in our experiment. This duration represents the lower time scale for the spontaneous symmetry breaking into structurally equivalent monoclinic lattices, whose presence possibly stabilizes the transient higher symmetry phase. Above the fluence threshold, the coherence length enhancement and the HT phase fraction growth both saturate at 5 mJ/cm$^2$, indicating that the transient non-equilibrium state is a precursor to the metallic state.

The size of the monoclinic crystallites $\xi$ we determine from X-ray nanoscopy is in excellent agreement with the electronic correlation length $\xi_{el}$ found in bulk V$_2$O$_3$ (Ref. [21]). We conclude a strong interdependence between the electronic correlations and structural twinning of the LT phase into different monoclinic crystallites in equilibrium. By assuming that the electronic correlation length and the crystallite size remain coupled after photoexcitation $\xi_{el}(\tau)=\xi(\tau)$, we estimate the behavior of the dynamic Mott gap via $\Delta(\tau)\sim 1/\xi(\tau)^2$ (Refs. [1,21]). The photoexcitation induces an ultrafast alignment of the crystallites, which effectively increases the coherence length and induces a narrowing of the band gap (see Fig. 3d). The correlation length increases by a factor of 1.3 at a pump fluence of 5 mJ/cm$^2$, thus the gap reduces from 0.5 eV in the ground state[22] below 0.3 eV in the photoexcited state. The maximum dynamic correlation length in the monoclinic phase is smaller than the correlation length in the rhombohedral phase; thus the gap is not closed entirely, consistent with the existence of a dynamic pseudogap observed in other Mott systems[23].

The photoexcited phase is structurally similar to the paramagnetic insulating phase, which in equilibrium occurs with chromium doping or equivalently with negative pressure[11–13]. Thus, the observation of the ultrafast quench of the monoclinic distortion indicates an ultrafast quench of the antiferromagnetic order. This is further supported by the negative thermal expansion in V$_2$O$_3$ within the hexagon plane, which results in transient negative pressure after photoexcitation: in equilibrium, the negative pressure reduces the antiferromagnetic order and the monoclinic distortion[13]. The photoinduced insulator-to-metal transition in $V_2O_3$ lasts for tens of picoseconds[7], consistent with the duration of the structural unit cell volume change reported here. The long-range order enhancement through cooperative nanoscale alignment occurs before the electronic transition completes. Because long-range order is essential for electronic transport[8], we anticipate that the presence of the photoinduced structural order impacts the dynamics of the electronic transition.

# Figures

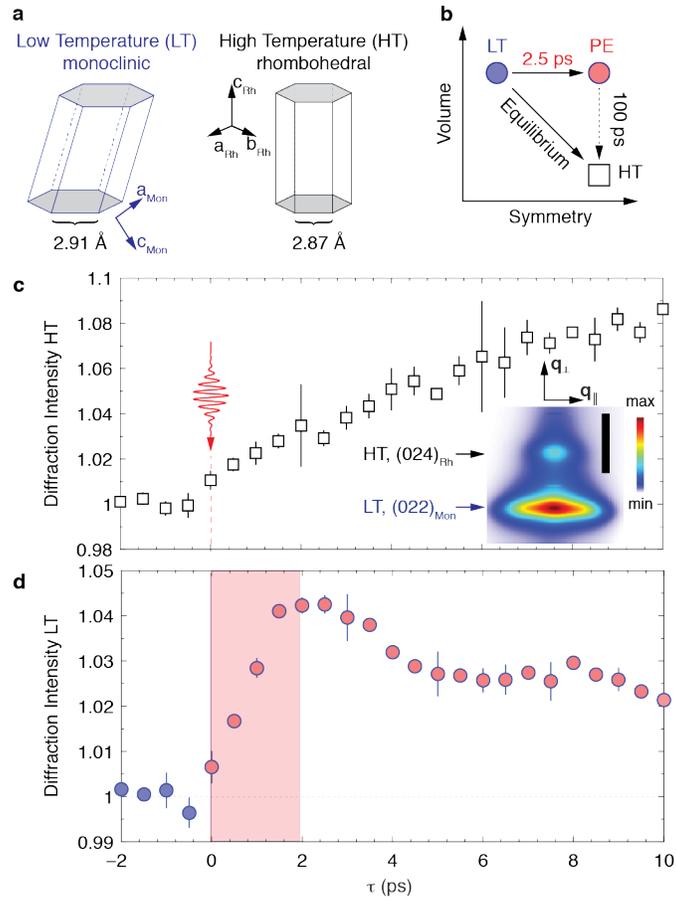

**Figure 1: Temperature induced and photoinduced structural phase transition**. **a,** The monoclinic insulating and rhombohedral metallic structures. The hexagon corners represent vanadium atoms. Oxygen and other vanadium atoms within the shown structures are omitted for better visibility. **b,** Schematic phase diagram of two equilibrium low-temperature (LT) and high-temperature (HT) phases and the photoexcited (PE) phase. In equilibrium, the transition occurs directly from LT to HT, while photoinduced transition occurs via a transient, non-equilibrium phase. **c, d,** The peak intensity of the HT, $(024)_{Rh}$ diffraction peak (**c**) and of the LT, $(022)_{Mon}$ diffraction peak (**d**) as a function of the time delay after photoexcitation (pump fluence 1 mJ/cm$^2$) at a base temperature of 157 K. Inset in **c**: A slice through the 3D diffracted intensity around both peaks. $q_\perp$ is parallel to $(024)_{Rh}$ and $q_\parallel$ to $(-210)_{Rh}$. Scale bar shows 0.04 Å$^{-1}$. The uncertainties in **c, d** result from the difference between two independent measurements.

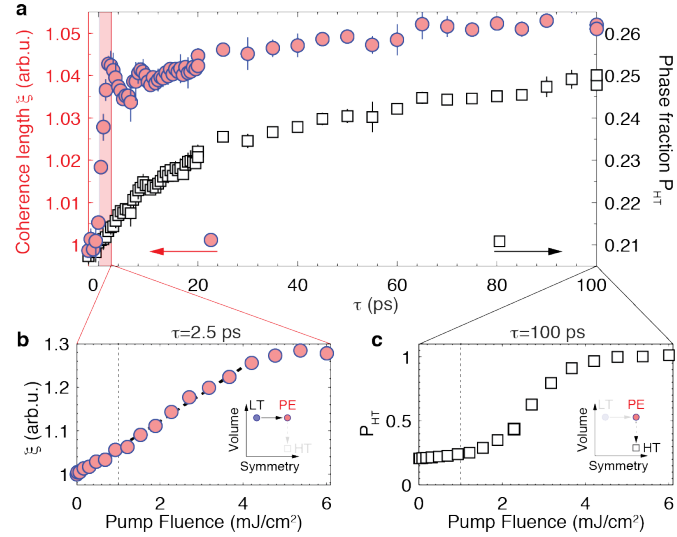

**Figure 2: Fluence dependence of the photoinduced structural phase transition. a,** Time dependence of the coherence length ξ (filled red circles, left axis) calculated as the reciprocal peak width with a fluence of 1 mJ/cm$^2$ and the HT phase fraction $P_{HT}=V_{HT}/(V_{HT}+V_{LT})$ (open black squares, right axis), where $V_{HT}$ and $V_{LT}$ are the HT and LT peak integral intensities. **b,** The coherence length 2.5 ps after photoexcitation as a function of the laser pump fluence. **c,** The HT phase fraction 100 ps after photoexcitation normalized to the value at highest fluences, where the entire film is metallic. Insets show the schematic path from LT to photoexcited (PE) phase **b** and PE to HT phase **c**.

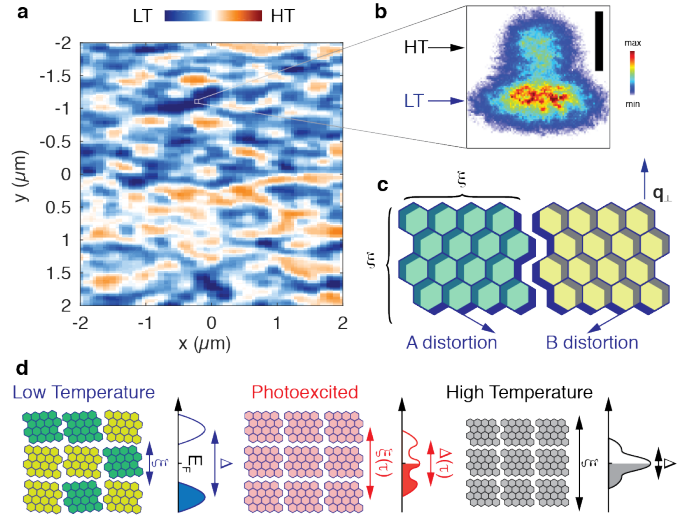

**Figure 3: Nanoscale description of the dynamic phase transition. a,** Measured spatial distribution of the LT phase in equilibrium determined from the integral intensity of the LT diffraction peak, $I_{LT}(x,y)$. The spatial distributions of the HT phase ($I_{HT}(x,y)$, not shown) is anti-correlated to the LT phase distribution with a Pearson correlation coefficient of $\frac{\int I_{LT}(x,y) \cdot I_{HT}(x,y) dxdy}{\sqrt{(\int I_{LT}(x,y)^2 dxdy)(\int I_{HT}(x,y)^2 dxdy)}} = -0.58 \pm 0.01$ (complete segregation yields a value of -1 and complete intermixing a value of 0). **b** A typical diffraction pattern measured during a single exposure to the nanobeam with a spot size of 20 nm. Scale bar shows 0.04 Å$^{-1}$. **c** Schematic of two neighboring monoclinic domains with distortions along different hexagon edges A and B. The monoclinic distortion and the physical touching of the mosaic blocks results in an angular misalignment. The size of the monoclinic domains is $\xi$=8 nm. **d** Schematic evolution of the structure and the electronic properties during the photoinduced phase transition. (Left panel) In the low temperature ground state the crystallites are misaligned and the electronic system is gapped. (Middle panel) Photoexcitation raises the coherence length through alignment of the crystallites and partially closes the Mott gap. (Right panel) The volume changes discontinuously, the Mott gap closes and the equilibrium metallic state emerges.

**Acknowledgements**

We thank Lu Sham, Marcelo Rozenberg, and Richard Averitt for discussions, and thank Eric Fullerton for pointing out the significance of the negative thermal expansion during the phase transition. The work at UCSD was supported by the AFOSR grant #FA9550-16-1-0026 and a UC collaborative grant MRPI MR-15-328-528. The work at UCSD was supported by U.S. Department of Energy, Office of Science, Office of Basic Energy Sciences, under Contracts No. DE - SC0001805 (x-ray scattering A.S., D.C., N.H., R.K., O.K., J.W., and O.G.S.) and No. DE FG02 87ER-45332 (thin film synthesis and characterization J.G.R., I.V., and I.K.S.). J.G.R. acknowledges support from FAPA program through Facultad de Ciencias and Vicerrectoria de Investigaciones of Universidad de los Andes, Bogotá Colombia and Colciencias #120471250659 and #120424054303. Use of the Linac Coherent Light Source (LCLS), SLAC National Accelerator Laboratory, is supported by the U.S. Department of Energy, Office of Science, Office of Basic Energy Sciences under Contract No. DE-AC02-76SF00515. Use of the Advanced Photon Source and the Center for Nanoscale Materials, both Office of Science User Facilities, was supported by the US Department of Energy, Office of Science, Office of Basic Energy Sciences, under Contract No. DE-AC02-06CH11357

**Author contributions**

A.S and J.G.R. conceived of the idea. A.S., J.G.R., I.K.S., and O.G.S planned the experiment. J.G.R. and I.V. prepared the samples, characterized them and performed resistance measurements for the in-situ temperature calibration. A.S, J.G.R, I.V., D.C, N.H., R.K., O.K., J.M.G., M.S., M.Ch., and O.G.S. conducted the time resolved experiment. A.S., N.H., J.W., and M.H. conducted the nanodiffraction experiment. A.S. analyzed the data. All authors discussed the interpretation of the data. A.S., J.G.R., I.V., and D.C. wrote the initial version of the paper. The paper underwent multiple revisions by all authors.

# Non-equilibrium phase precursors to the insulator-metal transition in $V_2O_3$
## Supplementary Information

Andrej Singer[1,*], Juan Gabriel Ramirez[1,2], Ilya Valmianski[1], Devin Cela[1], Nelson Hua[1], Roopali Kukreja[1], James Wingert[1], Olesya Kovalchuk[1], James M. Glownia[3], Marcin Sikiroski[3], Matthieu Chollet[3], Martin Holt[4], Ivan K. Schuller[1], and Oleg G. Shpyrko[1]

[1]Department of Physics and Center for Advanced Nanoscience, University of California San Diego, La Jolla, California 92093, USA

[2] Department of Physics, Universidad de los Andes, Bogotá 111711, Colombia

[3]LCLS, SLAC National Accelerator Laboratory, Menlo Park, California 94025, USA

[4]Center for Nanoscale Materials, Argonne National Laboratory, Argonne, IL 60439, USA

[*]Present address: Department of Materials Science and Engineering, Cornell University, Ithaca, NY, 14850, USA

**Experiment description**

We deposit $V_2O_3$ thin films on R-plane sapphire substrates by RF magnetron sputtering[1]. The film thickness is 100 nm, determined by laboratory x-ray reflectivity measurements. Transport measurement show a resistance change of 5 orders of magnitude (see Fig. S1). The time-resolved x-ray experiment is conducted at the XPP instrument of the LCLS. To control the temperature, we use a cryo-jet and conduct the experiment at ambient conditions. An x-ray spot of 200 μm x 200 μm, a laser spot of 500 μm x 500 μm, and collinear x-ray and optical laser pulses are used. The nanodiffraction data is collected at the ID-26 instrument at the Argonne Photon Source; the temperature is controlled by a liquid nitrogen flow cryostat and the experiment is done in vacuum. All experiments presented in the main text and the supplement are conducted on a film grown on a single substrate piece, sliced into smaller pieces for further measurements.

**Recording and analyzing the 4D dataset**

We acquire the 3D reciprocal space representation by rocking the crystal from -1 to +1 degrees in 21 steps and by transforming the data into a Cartesian grid in reciprocal space, similar to the work in Bragg Coherent Diffractive Imaging[2]. A representation of the 4D dataset is shown in Supplementary Figure S3. The inset in Figure 1c of the main text shows a projection along the normal to patterns shown in Figure S3 (**a, e, i**). Figures 1c,d show the results by directly analyzing the height both peaks in these images, which were measure for time delays from -2 to 10 ps. For longer time delays shown in Figures 2**a** we first measure a single slice through the reciprocal space (the geometrical direction of this slice is indicated in Figures S3 **b, c**), and second, calculated the peak width from the peak height in this slice assuming the total intensity of the whole 3D reciprocal space remains constant during the dynamic structural phase transition. The latter analysis is in agreement with the direct calculation of the width and height of the peaks for time delays from -2 to 10 ps.

**Diffraction at higher q values**

The width of the low temperature peak shown in the inset of Figure 1**c** can be well approximated with three Gaussian functions: two symmetric side Gaussians and a Gaussian in the center. To understand the nature of these three contributions, we collect x-ray diffraction data at high momentum transfer q along $q_\perp$, i.e. $(022)_{Mon}$ (data in the main text), $(033)_{Mon}$, $(044)_{Mon}$ (see Fig. S6). The peak in the center significantly broadens at high q, indicating high tilt distribution in the structure corresponding to this peak, while the side peaks retain their widts. This observation is similar to the Williamson-Hall analysis[3]. However, here we look perpendicular to the Bragg peak and the strain in the Williamson-Hall analysis is consistent with the 'tilt' of the Bragg planes[4].

**Statistical analysis of the nanodiffraction data**

To estimate the number of mosaic blocks in the volume exposed to the x-ray beam, we assume each block is either in orientation A or B. (The peak corresponding to the third monoclinic distortion is significantly suppressed in diffraction and is only noticeable at lowest measured temperatures of 130 K.) Assuming we illuminate N blocks with the nanobeam the total x-ray intensity is calculated via $I = P \cdot y_A + N/2 \cdot y_S + (N-P) \cdot y_B$, where $y_A$ and $y_B$ are the intensities of a single block in orientation A and B, and $y_S$ is a strain term which does not depend on orientation (see diffraction at higher **q** in Fig. S6). P is the number of blocks in orientation A calculated assuming a binomial distribution with a total number of blocks N and the probability of 0.5. A total of 20000 intensity profiles $I_j$ are calculated for a number of mosaic blocks N, j=1…N. Subsequently, the ratio of the width of the average intensity (width of $I_{avg}=\Sigma I_j$) to the average of the widths of single intensities is determined. This ratio in the data is 1.0035, which according to our analysis corresponds to N=200 of mosaic blocks in the beam volume. Given the volume of the beam 20 nm·20 nm ·100 nm /sin(23°)=102000 nm$^3$ yields an average volume of around 500 nm$^3$ per cube. Assuming cubic blocks we conclude that each block is (8 nm)$^3$ in size.

# Figures

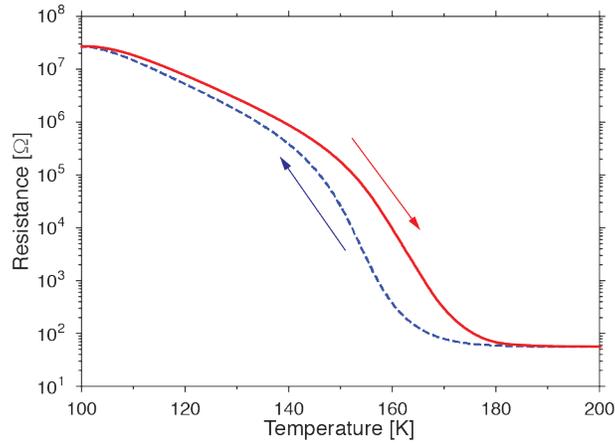

**Figure S1: Transport measurements.** The resistance measured on the film used for time resolved and spatially resolved experiments and used to calibrate the temperature. Red solid and blue dashed lines show the heating and cooling branches.

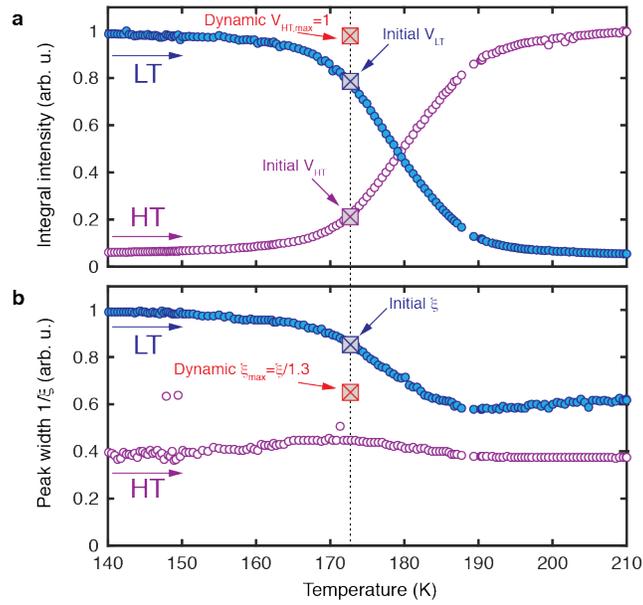

**Figure S2: Static phase transition.** Quasi-static insulator-to-metal transition measured with a macroscopic (spot size ~ 200 µm), parallel beam at the APS synchrotron source. **a** Phase fraction of the monoclinic $P_{LT}=V_{LT}/(V_{LT}+V_{HT})$ (blue filled circles) and rhombohedral $P_{HT}=V_{HT}/(V_{LT}+V_{HT})$ (open magenta circles) phases determined from the integral peak intensity, averaged in the direction perpendicular to the scattering vector $q_\perp$. $V_{LT}$ and $V_{HT}$ are the integral peak intensities. The ground state in time resolved experiments is indicated by arrows. **b** The peak width in the direction $q_\parallel$ $(-210)_{Rh}$ perpendicular to the scattering vector $q_\perp$

(also the surface normal). The 3D reciprocal space is projected onto the plane defined by $q_\perp$, $q_\parallel$, and the widths are determined as root mean squares. The LT peak width in the time resolved experiment is shown for the ground state, and the maximum correlation length enhancement at $\tau=2.5$ ps and a fluence of 5 mJ/cm$^2$. The smallest peak width in the time resolved experiments (highest correlation length) is close to the minimum peak width measured in equilibrium at 190 K.

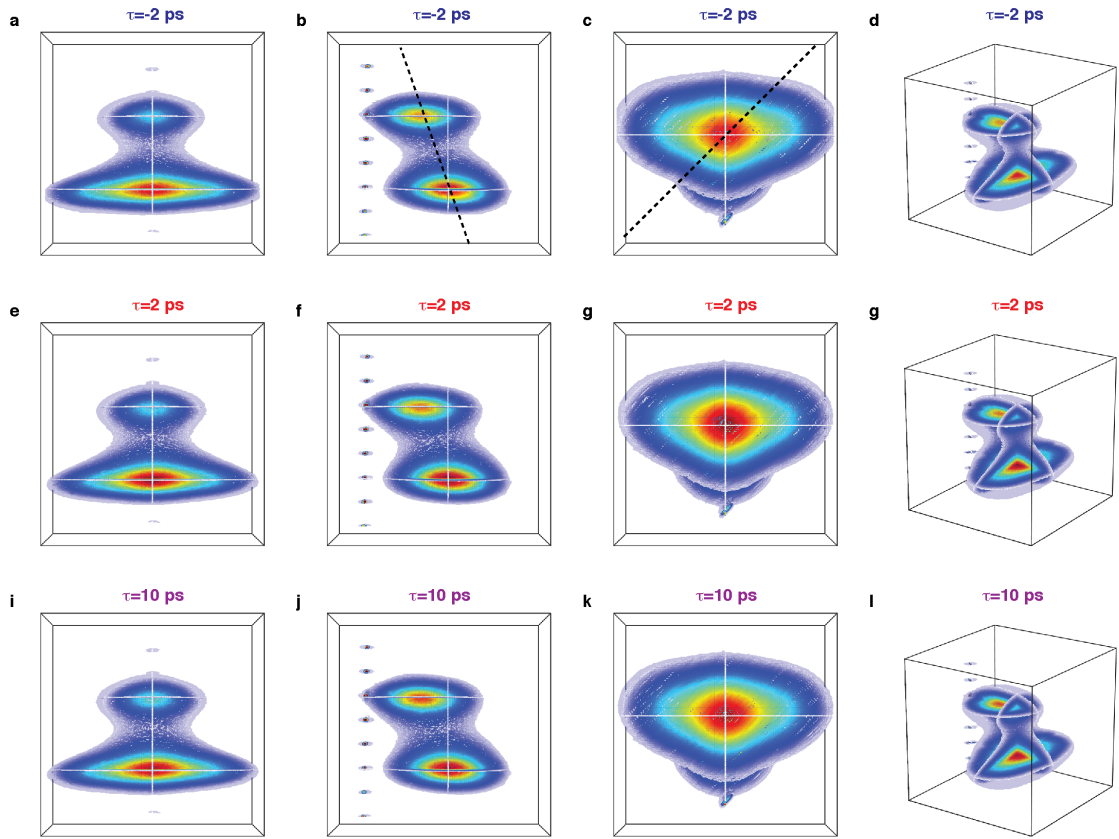

**Figure S3: Analysis of the 4D dataset.** 3D reciprocal space measured in the ground state (**a-d**), and at time delays of 2 ps (**e-g**) and 10 ps (**i-l**). The data was measured at a base temperature of 157 K and images represent different views on the 3D reciprocal space. The vertically arranged dots visible best in (**b, f, j**) are due to the substrate scattering. Note the reduction of the width of the monoclinic peak (lower peak) in (**a, e, i**) and the growth of the rhombohedra peak in **j**. The intersection with the Ewald sphere is shown in **b, c**. The separation between peaks is 0.04 Å$^{-1}$ and the angular width of the LT peak (broader peak) is 1° in (**a, e, i**) (compare with Fig. S6).

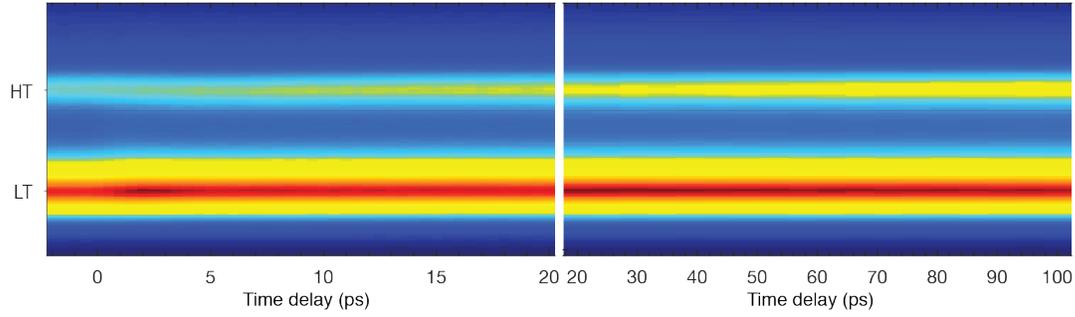

**Figure S4**: **Reciprocal space as a function of time delay for long time delays.** The HT and LT Bragg peaks are visible in a single slice by the Ewald sphere through the reciprocal space (the position of the slice is indicated by dashed lines in Fig. S3. The Ewald sphere is approximated flat in the angular region.). The LT peak height grows within 2.5 ps and remains high, while the HT peak height grows monotonically. A single slice was used for the results shown in Fig. 2a. The width of the LT peak was determined from its height (visible here) and the conservation of integral intensity of the sum of both peaks. The width of the HT peak does not change.

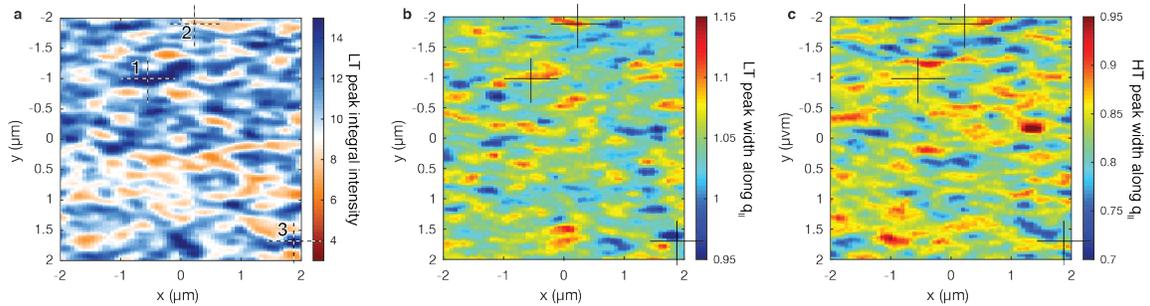

**Figure S5**: **The peak width in the nanodiffraction experiment. a** Figure 3 **a** shown for comparison. **b** the width of the LT peak and **c** the width of the HT peak along the $q_{\parallel}$ direction (determined as root mean square values from horizontal line scans through both peaks in images similar to Fig. 3b). The LT peak is wider than the HT peak, however, the difference in widths is slightly smaller than in the time resolved measurements, because the angular divergence of the nanobeam widens both peaks. No correlations between the LT peak intensity and LT peak width (inverse coherence length) is observed. For example, point 1 (indicated by crosses in **a, b, c**) has high peak intensity and a large width, point 2 (indicated by crosses in **a, b, c**) has a low peak intensity and a large width, point 3 (indicated in **a, b, c**) has a large peak intensity and a is in between regions with large and small width.

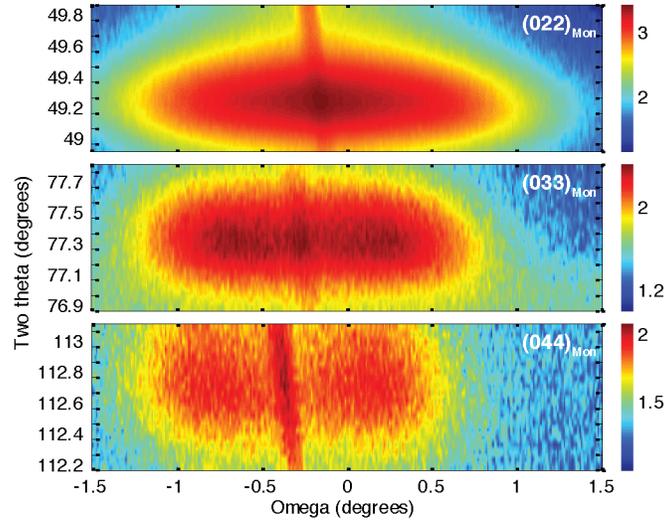

**Figure S6: Measurements at higher momentum transfers.** Rocking scans at higher q (two theta) values recorded with an x-ray tube (Cu) at a base temperature of 100 K. The central peak broadens and disappears for higher $(033)_{Mon}$ and $(044)_{Mon}$ reflections. The sharp line in the center of the pattern is absent in the free electron laser or at the synchrotron measurements and presumably results from substrate scattering of secondary lines in the x-ray tube spectrum (no monochromator is used due to the low signal to noise ratio).

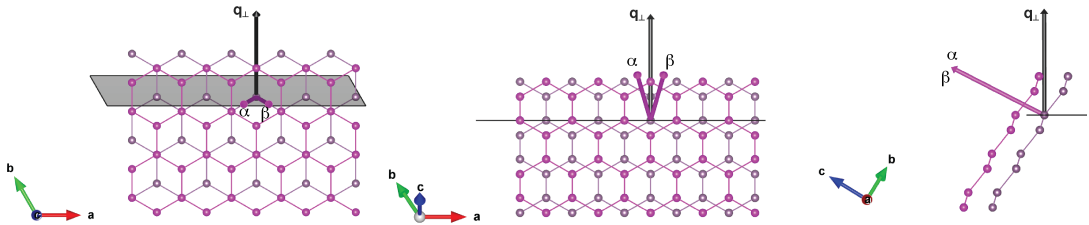

**Figure S7: Microscopic model for the twinning of the LT peak in thin films grown along $(024)_{Rh}$ direction.** Three views of the 3D rendering of the structure. $q_\perp$ denotes the direction of the out of plane $(022)_{Mon}$ Bragg peak. The vectors $\alpha$ and $\beta$ show the direction of the distorted c-axis in the rhombohedral system for two equivalent twins described in the main text Fig. 3c by A and B. The distortion represents a rotation of the c-axis onto $\alpha$ or $\beta$. The same rotation applied to the $q_\perp$ vector results in splitting of the $q_\perp$ vector into $q_1$ and $q_2$, which are split along the $(-210)_{Rh}$ direction and in the full splitting case result in 1.6°. In our x-ray data at 100 K, we observe a splitting in the same direction by 1°, which supports our microscopic twinning model and indicates a strain region between the twins. The splitting angle is lower at higher temperatures (see Fig. S2).

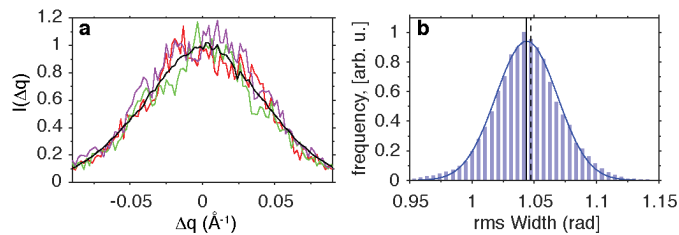

**Figure S8: Statistical analysis of the nanodiffraction data. a** Typical line scans through the LT bragg peaks in the diffraction patterns similar to Fig. 3b, measured with the nanobeam at different sample locations (colored shaded lines) and the line scan from the average over all 20000 diffraction patterns measured at different locations (black solid line). All curves are normalized to one and the width of the single lines is similar to the width of the average. **b** A histogram of the root mean square widths calculated from the ensemble of all measured locations (blue bars). The mean of the histogram distribution is determined by a Gaussian fit (blue line) and is shown by the vertical black solid line. The vertical dashed line shows the width of the line scan of the average over all 20000 patterns (shown by the black line in **a**).

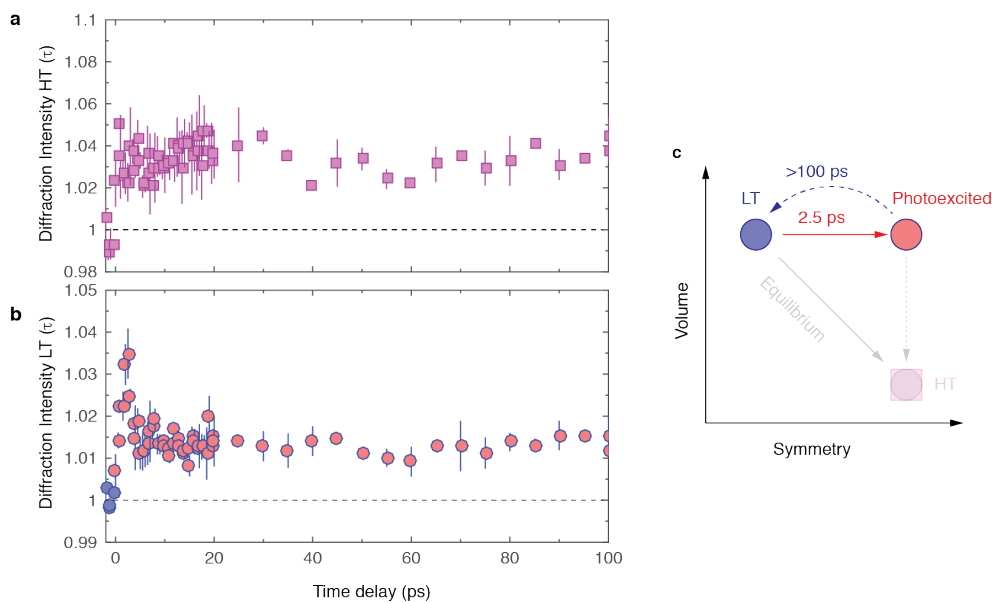

**Figure S9**: **Pumping below the fluence threshold. a, b** High temperature **a** and low temperature **b** peak response when pumped below the fluence threshold at a base temperature of 153 K. **c,** The corresponding schematic phase diagram. The long 100 ps time scale in **a** is absent in the high temperature peak response. The ultrafast increase of the HT diffraction peak is associated with the increase of the LT peak, whose tail has a non-vanishing intensity at the position of the HT peak.